\documentclass[aps,preprint,showpacs,preprintnumbers,amsmath,amssymb]{revtex4}
\usepackage{amssymb}
\usepackage[T1]{fontenc}
\usepackage[latin1]{inputenc}
\usepackage{graphicx}
\usepackage{dcolumn}
\usepackage{bm}

\begin{document}

\title{Ferromagnetic instabilities in neutron matter at finite
       temperature with the Gogny interaction}

\author{D. L\'opez-Val, A. Rios, A. Polls and I. Vida\~na}

\address{Departament d'Estructura i Constituents de la Mat\`eria,
         Universitat de Barcelona, Avda. Diagonal 647, E-08028 Barcelona, Spain}

\begin{abstract}
The properties of spin polarized neutron matter are studied both at zero and
finite temperature using the D1 and the D1P parameterizations of the Gogny
interaction. 
The results show two
different behaviors:
whereas the D1P force exhibits a ferromagnetic transition at a density of
$\rho_c \sim 1.31$ fm$^{-3}$ whose onset increases with temperature,
no sign of such a transition is found for D1 at any
density and temperature, in agreement with recent microscopic calculations.
\end{abstract}

\vspace{0.5cm}

\maketitle

The possible existence of a phase transition of neutron matter to a
ferromagnetic state has motivated many investigations of the Equation of
State (EoS) of spin-polarized neutron  matter.
Besides the interest that such a transition could have in the context of neutron
stars \cite{pra97}, this problem has gained interest in itself 
and has been addressed in the framework of very different theoretical
approaches \cite{ha75,da78,vi84,ku94,be95,fa01,ma02,vi02a,vi02b,is04a}.
Whereas some of these calculations, like for instance those based on Skyrme-like
interactions, predict a transition at densities in the range $(1-4)\rho_0$
(with $\rho_0=0.16$ fm$^{-3}$ the saturation density of symmetric nuclear
matter), others, like recent Monte Carlo \cite{fa01} or Brueckner--Hartree--Fock (BHF)
calculations \cite{vi02a,vi02b} using modern two- and three-body realistic
interactions, exclude such a transition at least up to densities around five
times $\rho_0$. In spite of this discrepancy, it is interesting to study how
temperature  influences the ferromagnetic
transition \cite{ri05,bo06}. 
In Ref.~\cite{ri05} an
analysis of the temperature effects in the framework of a Hartree-Fock
calculation with Skyrme interactions was performed. In the present paper, we would like to
study the influence  of the finite-range terms of the interaction
on the ferromagnetic transition.
To this end, we will consider the Gogny interaction. This is
an effective nucleon-nucleon force with both zero- and
finite-range terms and a simple spin-isospin structure.
In addition to the original D1 parameterization \cite{go80},
several other parameterizations of this force have been obtained.
The D1S force, for instance, was introduced to improve
the pairing properties and surface effects of finite nuclei~\cite{be91}, while
more recently the D1P \cite{fa99} interaction has been introduced with the aim
of reproducing the EoS of pure neutron matter given by a variational microscopic
calculation with realistic interactions \cite{wi88}.

The properties of nuclear matter deduced from Gogny interactions
have already been treated in the literature
\cite{zha96}.
Indeed, several instabilities produced by
these forces at zero temperature have
been studied in previous works \cite{is04b}. The isospin instability, for instance,
is a common feature to all the existing Gogny parameterizations as well as of most Skyrme
forces. 
This instability is signaled by the fact that,
above a certain critical density, the energy per particle of nuclear matter becomes
more repulsive than that of neutron matter.
For the D1P force, this instability takes place at
$\rho_I \sim 7 \rho_0$, while for D1
it occurs at $\rho_I \sim 3 \rho_0$. Furthermore, D1 and D1S
also exhibit a spinodal instability in neutron matter, \emph{i.e.}, the energy per
particle does not increase monotonically with
density. Instead, it reaches a maximum value at a critical
density (which for D1 and D1S is around $\rho_S \sim 4\rho_0$) and
then decreases from that density on.
For the discussion of our results, we will choose the D1P and D1
parameterizations of the Gogny NN force, since they represent
two qualitatively different behaviors regarding ferromagnetic
instabilities.


In general, any Gogny interaction can be casted in the following form:
\begin{eqnarray}
V_{NN}({\vec r})&=&
\sum_{i=1}^{2} e^{-\frac{r^2}{\mu_i}}
\Big( W_i+B_i\,P_{\sigma}-H_i\,P_{\tau}-M_i\,P_{\sigma}\,P_{\tau} \Big)
+ t_0\,\Big(1+x_0\,P_{\sigma}\Big)\rho_N^{\alpha}\,\delta ({\vec r})
\label{eq:gogny_force}
\end{eqnarray}
with $\vec{r}$ the distance between two nucleons.
The spin-isospin structure of the force is given by the spin (isospin) exchange operators
$P_{\sigma}$ ($P_{\tau}$). Notice that the usual spin-orbit term of the
Gogny interaction has been omitted, because it does not give any contribution in
infinite matter. The Gogny force includes a sum of two gaussian-shaped terms
which mimic the finite-range effects of a realistic interaction in the medium.
Usually, it also contains one density-dependent zero-range term, even though in the case of the
D1P parameterization two of these terms were used to make the
fitting procedure more flexible.

In the following, we will consider spin-polarized neutron matter,
characterized by the spin polarization parameter 
$\Delta=(\rho_\uparrow - \rho_\downarrow)/(\rho_\uparrow + \rho_\downarrow)$,
where $\rho_{\uparrow (\downarrow)}$ is the density corresponding to neutrons having spin
up (down) respect to a given direction, and by the total density, $\rho=\rho_{\uparrow} + \rho_{\downarrow}$.
The energy per particle in the Hartree--Fock approximation is given by
\begin{eqnarray}
e(\rho_{\uparrow}, \rho_{\downarrow},T) &=& \frac {1}{\rho} \sum_{\sigma,k}
 \frac {\hbar^2 k^2}{2m} \, n_{\sigma}(k,T)
\nonumber \\
&+& \frac{1}{2 \rho} \sum_{\sigma_1, k_1; \sigma_2, k_2} \langle {\vec k_1} \sigma_1, {\vec k_2} \sigma_2 \mid V_{NN} \mid {\vec k_1} \sigma_1, {\vec k_2} \sigma_2 \rangle_A \times
n_{\sigma_1}(k_1,T) \, n_{\sigma_2}(k_2,T) \, ,
\label{eq:energy}
\end{eqnarray}
where $n_{\sigma}(k,T)$ are the momentum distributions:
\begin{eqnarray}
n_{\sigma}(k,T) = \frac{1}{1 + e^{\beta [ \epsilon_{\sigma}(k) - \mu_{\sigma}] }} \, 
\label{eq:momdis}
\end{eqnarray}
of each spin component $\sigma$.
At each temperature $T$ and density $\rho_\sigma$, a self-consistent procedure has to be performed to compute the
chemical potential of each species, $\mu_{\sigma}$, from the normalization condition:
\begin{eqnarray}
\rho_{\sigma} = \sum_k n_{\sigma}(k,T) \, .
\label{eq:fd}
\end{eqnarray}
Once the momentum distribution is determined, the entropy per particle of the system is given by:
\begin{eqnarray}
s(\rho_{\uparrow},\rho_{\downarrow},T) &=&
\frac {1}{\rho} \sum_{\sigma,k} \, \big\{ n_{\sigma}(k,T) \ln n_{\sigma}(k,T) \nonumber \\
&+& [1- n_{\sigma}(k,T)] \ln[ 1- n_{\sigma}(k,T) ] \big\} \, .
\label{eq:entro}
\end{eqnarray}
From the internal energy and the entropy one can readily compute the free energy
per particle,
$ f(\rho_{\uparrow},\rho_{\downarrow},T ) =
e(\rho_{\uparrow},\rho_{\downarrow},T)- T
s(\rho_{\uparrow},\rho_{\downarrow}, T)$,
and, from this, the inverse magnetic susceptibility is given by:
\begin{equation}
\frac{1}{\chi}=\frac{1}{\mu^2 \rho}
\left( \frac{\partial^2 f}{\partial \Delta^2} \right)_{\Delta=0} \; ,
\label{eq:mag1}
\end{equation}
with $\mu$, the magnetic moment of the neutron.

An important quantity for our analysis is the single-particle (sp) energy:
\begin{eqnarray}
\varepsilon_\sigma(k) = \frac{\delta e}{\delta n_\sigma(k)}=
\frac{\hbar^2k^2}{2m}+U_\sigma(k) \,
\end{eqnarray}
where $U_\sigma(k)$ is the sp potential. Note that in contrast to the
sp spectrum $\epsilon_\sigma(k)$ appearing in Eq.~(\ref{eq:momdis}), this sp
energy contains the rearrangement effects.
The momentum dependence of the sp potentials can be characterized
by the effective mass $m^*_\sigma(k)$:
\begin{eqnarray}
\frac{m^*_\sigma(k)}{m}=\left. \left[ \frac{m}{\hbar^2 k'} \frac{d
\varepsilon_\sigma}{dk'} \right]^{-1} \right|_{k'=k}
= \left. \left[ 1 + \frac{m}{\hbar^2 k'} \frac{d U_\sigma}{dk'} \right]^{-1} \right|_{k'=k} \, .
\label{eq:mass}
\end{eqnarray}

\begin{figure}[t]
   \includegraphics[width=0.75\textwidth]{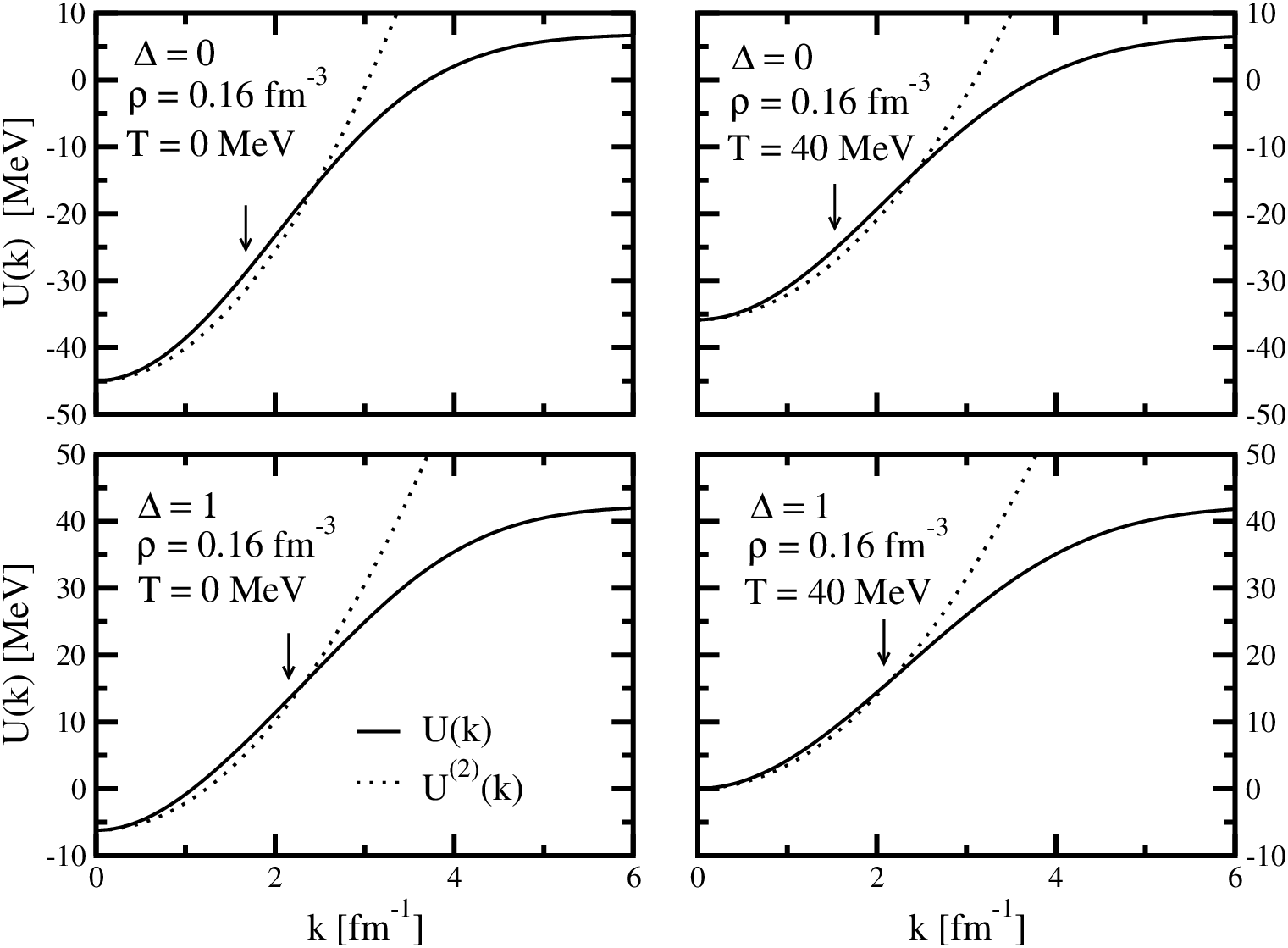}
   \vspace{0.75cm}
   \caption{Single-particle potential $U(k)$ 
for up neutrons in non-polarized (top panels) and totally polarized (bottom
panels) neutron matter at 
 $\rho=0.16$ fm$^{-3}$ for $T=0$ (left panels) and $T=40$ MeV (right panels) within the
D1P parameterization. $U(k)$ and its quadratic approximation $U^{(2)}$ are displayed
in full and dashed lines respectively. The arrows denote the value of $k_F^{\uparrow}$.}
   \label{fig:fig1}
\end{figure}

Let us start our analysis with the properties of the sp potential, $U(k)$.
Fig.~\ref{fig:fig1} reports the sp potential obtained from D1P for
neutrons with spin up at saturation density in a non-polarized
system ($\Delta=0$) and in a fully polarized one ($\Delta=1$) for  
T=0 (left panels) and 40 MeV (right panels).
The splitting due to the different spin polarizations can be understood
in terms of both the dependence of the sp potentials in phase space and the
dependence on spin of the effective NN interaction \cite{bo06}. Moreover, the
increase of temperature makes the sp potentials slightly less attractive in both
cases due to the dependence on the transferred momentum of the two-body matrix
element which, at finite temperature, is explored in a wider range of momenta.
In spite of the fact that the effective mass depends on momentum, for low
enough values of $k$ the sp potential can be casted in a quadratic form:
\begin{equation}
U^{(2)}_{\sigma}(k) = U_{\sigma}(0) + \left.
\left[\frac{1}{2k}\frac{dU_\sigma(k)}{dk}\right]\right| _{k^\sigma_F}k^2 \, ,
\label{eq:approx}
\end{equation}
in which the derivative $\frac{dU_\sigma(k)}{dk}$ and the $k$-independent term $U_\sigma(0)$ are determined from the full sp spectrum at the corresponding temperature.
Fig. \ref{fig:fig1} indeed illustrates that, for the density and polarization under
study, this quadratic approximation (dotted lines) is very close to the full $k$-dependent $U_\sigma(k)$ sp potential
(full lines), at least for momenta up to $k^\sigma_F$.
The approximation of Eq.~(\ref{eq:approx}) is devised to reproduce the slope of the spectrum at $k_F$ together with its value at $k=0$. The choice of the Fermi momentum $k_F^\sigma$ is related to the fact that the most relevant changes in the occupation of the sp states, when changing $\rho$ and $T$, essentially involve the neighboring region of the Fermi surface.
The quadratic approximation of Eq.~(\ref{eq:approx}) leads to an approximated form for the sp energies in terms of the effective mass at the Fermi surface and a k-independent term of $U^{(2)}_\sigma(k)$:
\begin{equation}
\epsilon_{\sigma}(k,T) \approx \frac{\hbar^2 k^2}{2m_{\sigma}^{*}(k_F^\sigma,T)} +
U_\sigma(k=0,T)\,.
\label{eq:eps}
\end{equation}
Finally, let us note that a similar picture is obtained for the other Gogny parameterizations.

\begin{figure}[t]
   \includegraphics[width=0.75\textwidth]{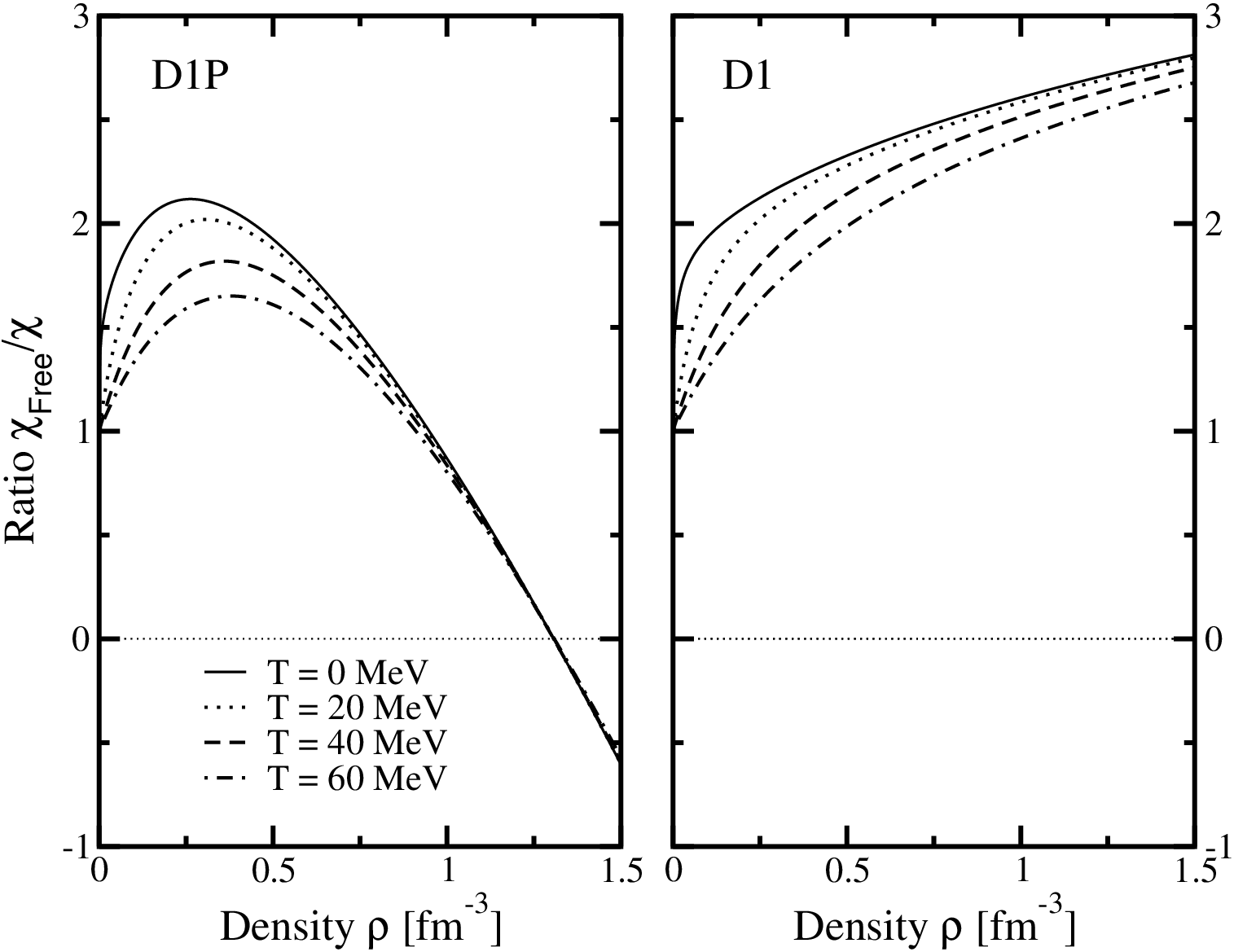}
   \vspace{0.75cm}
   \caption{Ratio between the inverse magnetic susceptibility of
interacting neutron matter and that of the corresponding free Fermi sea
as a function of density for several temperatures.}
\label{fig:fig2}
\end{figure}

In Fig.~\ref{fig:fig2} we report the ratio $\chi_{Free}/\chi$ (where
$\chi_{Free}$ is the magnetic susceptibility of the free Fermi sea at the
corresponding temperature) as a function of density. Results for the
D1P (D1) forces are shown on the left (right) panels of the figure at
several temperatures. A different qualitative behavior for
the two parametrizations is clearly observed.
On the one hand, the D1P parameterization
leads, similarly to what was found in the case of the Skyrme interaction
\cite{ri05}, to a ferromagnetic phase transition, signaled by a
vanishing ratio $\chi_{Free}/\chi$. 
The critical density of this transition at T=0 MeV is
$\rho_{c} = 1.31$ fm$^{-3}$, a much larger value than the ones
obtained with Skyrme interactions, all of which were systematically
below $1$ fm$^{-3}$ (see Tab. II of Ref.~\cite{ri05}).
Moreover, and even though it is difficult to distinguish this in the figure,
the critical density slightly increases with temperature. This
is in agreement with the intuitive idea that thermal disorder increases the onset
density of ferromagnetism, but it is opposite to what was found with Skyrme
forces.
On the other hand, no trace of a ferromagnetic transition is
seen for D1 at any density or temperature.
This behavior is very similar to what was found in a
BHF calculation with realistic interactions \cite{bo06}.
In addition, the temperature dependence (the higher the temperature, the lower the ratio becomes) is also in agreement with such microscopic calculations.

\begin{figure}[t]
   \includegraphics[width=0.75\textwidth]{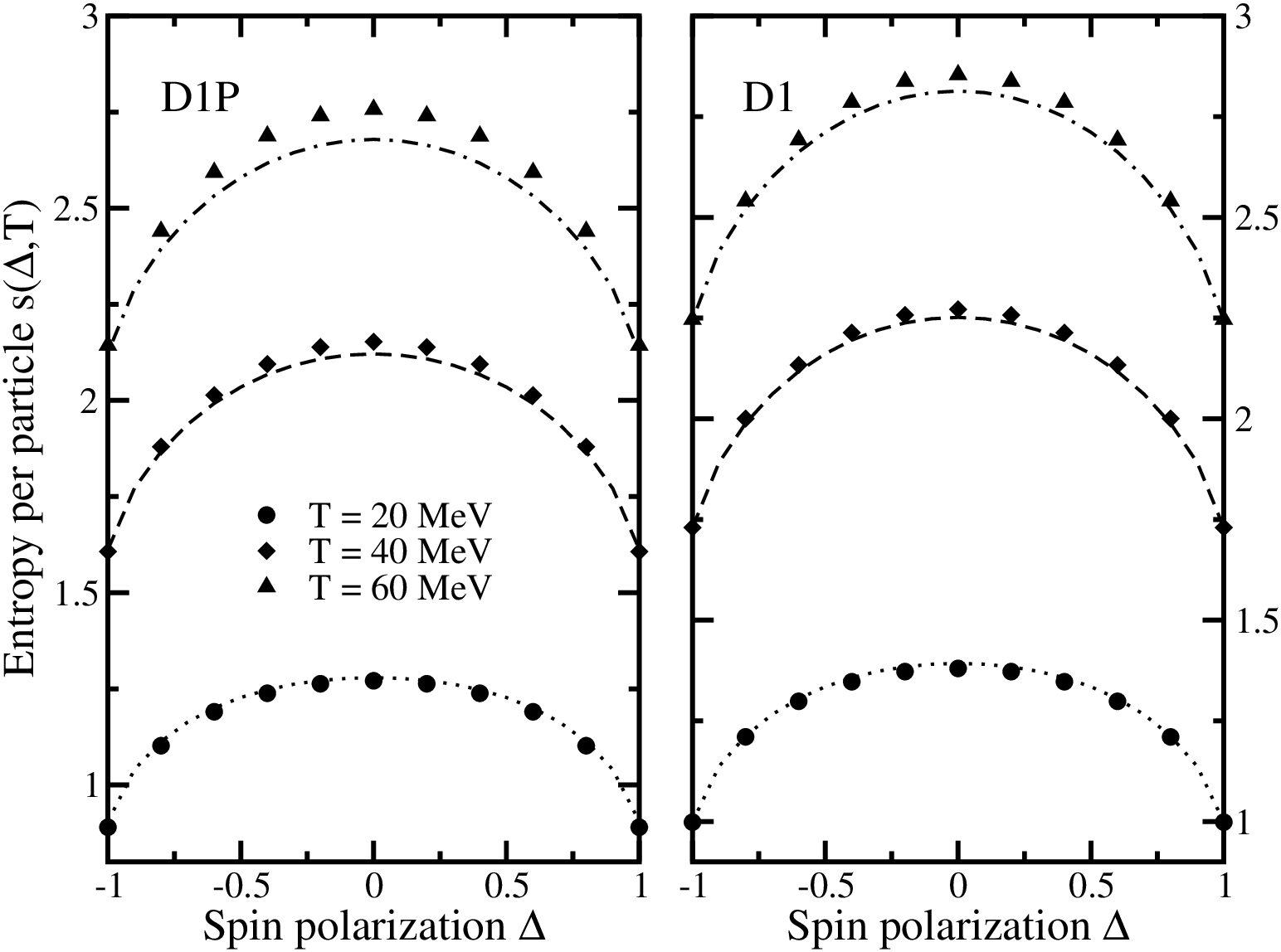}
   \vspace{0.75cm}
   \caption{Entropy per particle as a function of the spin polarization
at $\rho = 0.16$ fm$^{-3}$ and several temperatures. The exact Hartree-Fock values are depicted with circles, diamonds and triangles, whereas the lines illustrate the results obtained from a quadratic approximation of the single-particle spectrum (Eq. (\ref{eq:eps})).}
   \label{fig:fig3}
\end{figure}

In this context, it is interesting to study the behavior of the entropy.
To this end, plots of the entropy per particle as a function of spin polarization
at a fixed value of density ($\rho = \rho_0$) and several
temperatures are shown in Fig.~\ref{fig:fig3}.
We consider the entropy given by Eq.~(\ref{eq:entro}) computed with both i) the full
momentum-dependent spectrum $\epsilon_{\sigma}(k)$ (symbols), and
ii) the quadratic approximation to $\epsilon_{\sigma}(k)$ of Eq.~(\ref{eq:approx})
(lines). In the two cases the exact chemical potential determined from the normalization of the momentum distribution,
Eq.~(\ref{eq:fd}), are used. The agreement between the values of the entropy coming from the exact and the approximated sp energies is rather satisfactory, the most significant discrepancies not being larger than a $3 \%$.
Both the D1 and the D1P parametrizations give rise to similar results: the entropy per particle
is symmetric with respect to the non-polarized state and shows a
maximum at zero polarization. 
In addition, the values of the entropy increase
with temperature, which is again an indication that the entropy per
particle behaves as naively expected.
This so-considered
``natural'' behavior was also found in the BHF analysis of Ref.~\cite{bo06}.
In contrast, for Skyrme forces  
the entropy per particle of the polarized phase is seen to be higher
than the non-polarized one for a certain density on \cite{ri05}.
This defines
a kind of ``critical'' density, which is smaller than $\rho_0$ for most 
Skyrme parametrizations. 
Such a non-intuitive behavior of the entropy as a function of the polarization
can be related to the dependence of the entropy on the effective mass, and a
condition for the effective masses can then be derived:
\begin{equation}
\frac{m^*(\rho,\Delta=1)} {m^*(\rho,\Delta=0)} < 2^{2/3} \, ,
\label{eq:criter} 
\end{equation}
if the quantity $s(\rho,\Delta=1)-s(\rho,\Delta=0)$ has to be always
negative. Most of the Skyrme forces analyzed in Ref.~\cite{ri05} violate
this criterion and thus lead to an ``anomalous" temperature dependence for
the onset density of ferromagnetism. 

Notice, however, that for Skyrme forces the effective mass is momentum
and temperature independent. This is actually not the case of neutron matter
described by means of Gogny forces: the effective masses
do depend on both momentum and temperature. Nevertheless, as it has
been previously discussed, the sp spectrum and the entropy per particle are correctly
described by a quadratical momentum dependence, with an effective mass
calculated at $k = k_F^\sigma$ for each  temperature.
Within this approximation, one can proof that the criterion of Eq.~(\ref{eq:criter})
is still valid at each temperature.
We have checked that both the D1 and the D1P forces fulfill the criterion of Eq.~(\ref{eq:criter})
in a vast region of the $(T,\rho)$ parameter space, which includes both
the classical and degenerate limits. Thus, as it has been previously observed,
we do not expect that the ferromagnetic transition (only present for D1P) has an 
anomalous thermal behavior.

One of the questions that  needs for a closer insight is why
neutron matter described with Gogny forces does not present a ferromagnetic
instability, or why it is only present at very high densities. 
Intuitively, one expects that the instability will be related to the
zero-range term.
This can be understood just by taking into account that the pure
zero-range term only involves S-wave contributions and, therefore, due to
Pauli principle, it is not possible to have a couple of neutrons with the
same spin interacting through the contact term. Therefore there is no
contribution to the total energy per particle of fully polarized matter
from a zero-range term, whereas for non-polarized neutron matter this
term is both strongly density-dependent and repulsive, thus contributing
to the spin instability. In contrast, the behavior of finite-range terms,
both direct and exchange, has the opposite dependence with polarization: for
a given density, the higher the polarization, the higher these contributions
become. In other words, if only finite-range terms were considered, the
non-polarized system would be energetically favored. In addition, the
density dependence of these
contributions is softer than that of the zero-range term. Therefore, the
competition between zero and finite-range effects is resolved in favor
of the first one at high values of the density. From this, it is also obvious to
understand why no transition is found for the D1 and the D1S
parametrizations. In these two cases, the fitting parameters are such
that the zero-range term does not contribute for any polarization,
because $x_0=1$. The zero-range term is therefore
not active in neutron matter, whereas the direct and exchange parts of
the finite-range terms behave as mentioned, energetically
favoring the non-polarized case at all densities. In contrast, for the
D1P force $x_0 \neq 1$, and the zero-range term is active. As a consequence,
the instability arises once we come across a high enough critical density.


In this work, we have studied the properties of polarized neutron
matter with neutrons interacting through Gogny forces,
both at zero and finite temperature. The results show two different
qualitative behaviors for the two parametrizations under study.
On the one hand, the D1P exhibits a ferromagnetic transition at what
can be considered a very high density, $\rho_c \sim 1.31$ fm$^{-3}$. 
On the other hand, no sign of a ferromagnetic transition is found
for the D1 parameterization at any density or temperature.
The cause of these two different behaviors is related to
the zero-range term, which contributes for D1P, but not for D1.
Finally, we have checked that the temperature dependence of the ferromagnetic transition
with D1P is the one expected by intuition, in agreement with the fact that the criterion
of Eq.~(\ref{eq:criter})
for the effective masses is respected in a wide range of densities and temperatures.


\section*{Acknowledgments}

A. Rios acknowledges the support of DURSI and the European Social Funds.
D. L\'opez-Val acknowledges the support from a MEC Collaboration Fellowship (Spain).


\begin{thebibliography}{99}

\bibitem{pra97} M. Prakash, I. Bombaci, Manju Prakash, P.J. Ellis, J.M. Lattimer and R. Knorren,
Phys.  Rep.  {\bf 280}, 1 (1997).

\bibitem{ha75} P. Haensel,
Phys. Rev. C {\bf 11}, 1822 (1975).

\bibitem{da78} J. Dabrowski, W. Piechocki, J. Rozynek and P. Haensel,
Phys. Rev. C {\bf 17}, 1516 (1978).

\bibitem{vi84} A. Vidaurre, J. Navarro and J. Bernabeu,
Astron. Astrophys. {\bf 135}, 361 (1984).

\bibitem{ku94} M. Kutschera and W. W\'ojcik,
Phys. Lett. {\bf B 325}, 271 (1994).

\bibitem{be95} P. Bernardos, S. Marcos, R. Niembro and M. L. Quelle,
Phys. Lett {\bf B 356}, 175 (1995).

\bibitem{fa01} S. Fantoni, A. Sarsa and K. E. Schmidt,
Phys. Rev. Lett. {\bf 87}, 181101 (2001).

\bibitem{ma02} J. Margueron, J. Navarro and N. V. Giai,
Phys. Rev. C {\bf 66}, 014303 (2002).

\bibitem{vi02a} I. Vida\~na, A. Polls and A. Ramos,
Phys. Rev. C {\bf 65}, 035804 (2002).

\bibitem{vi02b} I. Vida\~na and I. Bombaci,
Phys. Rev. C {\bf 66}, 045801 (2002).

\bibitem{is04a} A. A. Isayev and J. Yang,
Phys. Rev. C {\bf 69}, 025801 (2004).

\bibitem{ri05} A. Rios, A. Polls and I. Vida\~na,
Phys. Rev. C  {\bf 71}, 055802 (2005).

\bibitem{bo06} I. Bombaci, A. Polls, A. Ramos, A. Rios and I. Vida\~na,
Phys. Lett. {\bf B 632}, 638 (2006).

\bibitem{go80} J. Decharg\'e and D. Gogny,
Phys. Rev. C {\bf 21}, 1568 (1980).

\bibitem{be91} J.F. Berger, M. Girod and D.Gogny,
Comp. Phys. Comm. {\bf 63}, 365 (1991).

\bibitem{fa99} M. Farine, D. Von-Eiff, P. Shuck, J.F. Berger, J. Decharg\'e and M. Girod,
J. Phys. {\bf G 25} (1999) 863.

\bibitem{wi88} R. B. Wiringa, V. Fiks and A. Fabrocini,
Phys. Rev. C {\bf 38}, 1010 (1988).

\bibitem{zha96} Y.-J. Zhang, R.-K. Su, H.Q. Song and F.-M. Li,
Phys. Rev. C {\bf 54}, 1137 (1996).

\bibitem{is04b} A. A. Isayev and J. Yang,
Phys. Rev. C {\bf 70}, 064310 (2004).

\end{thebibliography}
\end{document}